\documentclass[aps,pre,twocolumn,groupedaddress,showpacs]{revtex4}
\usepackage{CJK}

\usepackage{graphicx}

\usepackage{dcolumn}

\usepackage{bm}
\usepackage[normalem]{ulem}
\usepackage{color}

\usepackage{epstopdf}
\def\rf#1{(\ref{#1})}

\newcommand{\bew}{\begin{widetext}}
\newcommand{\nn}{\nonumber}
\newcommand{\ew}{\end{widetext}}

\newcommand{\bv}{\mathbf{v}}

\newcommand{\br}{\mathbf{r}}

\newcommand{\hp}{\mathbf{\hat{p}}}

\newcommand{\bs}{\mathbf{s}}

\newcommand{\bof}{\mathbf{f}}

\newcommand{\sep}{ \ \ \ , \ \ \ }

\newcommand{\beq}{\begin{equation}}
\newcommand{\eeq}{\end{equation}}
\newcommand{\beqn}{\begin{eqnarray}}
\newcommand{\eeqn}{\end{eqnarray}}
\newcommand{\pp}{\partial}
\newcommand{\dd}{{\rm d}}
\newcommand{\ee}{{\rm e}}

\newcommand{\la}{\langle}

\newcommand{\ra}{\rangle}

\newcommand{\vnab}{{\bf \nabla}}

\begin{document}

\begin{CJK*}{UTF8}{gbsn}

\title{
Birth, Death, and Horizontal Flight: Malthusian flocks with an easy plane in three dimensions}
\author{John Toner}
\email{jjt@uoregon.edu}
\affiliation{Department of Physics and Institute for Fundamental Science, University of Oregon, Eugene, OR $97403$}

\begin{abstract}
I formulate the theory of three dimensional ``Malthusian flocks" -- i.e., coherently moving collections of self-propelled entities (such as living
creatures) which are being ``born"  and ``dying" during their motion -- whose constituents all have a preference for having their velocity vectors lie parallel to the same two-dimensional plane. I determine the universal scaling exponents characterizing such systems exactly, finding that the dynamical exponent $z=3/2$, the ``anisotropy" exponent $\zeta=3/4$, and the ``roughness" exponent $\chi=-1/2$. I also give the scaling laws implied by these exponents.
\end{abstract}
\maketitle
\end{CJK*}

\section{Introduction and summary of results}{\label{Intro}}

In addition to its obvious importance to biology, the study of ``Active matter" \cite{book, Active1,Active2,Active3,Active4, tissue,PS,Nematics} has been of immense interest to physicists because of the novel collective phenomena that occur in such systems. These include long-ranged orientational order in spatial
dimension $d=2$ \cite{Vicsek,TT1,Chate1,Chate2}, and the breakdown of linearized hydrodynamics that occurs in
polar ordered active fluids ; aka ``flocks" \cite{TT1,TT3,birdrev}.

Reproduction (along with its inevitable companion, death) is another important non-equilibrium biological phenomenon.  One class of systems that exhibit both spontaneous motion and birth and death is
so-called  ``Malthusian flocks"\cite{toner_prl12, Malthus_1789, maltd>3prl, maltd>3pre}; that is,  polar ordered active fluids  in which the number of active particles is {\it not} conserved. Such systems are readily experimentally realizable in experiments on a, e.g.,
growing bacteria colonies and cell tissues\cite{tissue}, and ``treadmilling''
molecular motor propelled biological macromolecules in
a variety of intracellular structures, including the cytoskleton\cite{Prost_cyto},
and mitotic spindles\cite{Brugues}, in which molecules are
being created and destroyed as they move.

This prior work has treated Malthusian flocks whose underlying dynamics are completely isotropic. However, in many biological systems, anisotropy plays an important role. For example, birds are much more inclined to fly horizontally than vertically (for obvious evolutionary reasons!).
Many oceanic creatures, such as plankton\cite{plankton} preferentially swim horizontally as well.

Another system that would most definitely be described the theory presented here would be a collection of Janus particles\cite{janus}, with the top hemisphere heavier than the bottom. If one then coated the {\it front} of the particles with a catalyst like platinum, such particles would preferentially move in a horizontal plane.

In this paper, I consider the effects of such anisotropy on three-dimensional Malthusian flocks. That is, I consider a flock that fills a three dimensional space, whose constituents have a preference for moving parallel to an ``easy plane", which I take to be the $x$-$y$ plane. The ``hard`` direction orthogonal to this axis I will  call $r_h$.   (I refrain from the obvious choice of calling this direction $z$, because I am reserving $z$ for the ``dynamical exponent`` described below.) The two-component velocity field of the flock $\bv(\br,t)$ is a function of the three-dimensional Cartesian co-ordinate $\br=(x,y,r_h)$
as well as time $t$. 

I focus on what is known in the field of active matter as a ``dry`` system; that is, one without momentum conservation. In a three dimensional system as I am considering here, this would correspond to flockers moving through a fixed, frictional medium (e.g., worms crawling through soil).

Similarly to the isotropic theory of dry Malthusian flocks treated in \cite{toner_prl12, Malthus_1789, maltd>3prl, maltd>3pre},  the hydrodynamic theory I develop here  may also be viewed as  a generic non-equilibrium $3$-dimensional $2$-component spin model\cite{SRspin}  (i.e., a non-equilibrium ``XY`` model), with space-spin couplings that are invariant under simultaneous rotations of space and spin between the two components of a spin $\bs(\br,t)$ and the 
$(x,y)$ coordinates of the three dimensional space. In such a spin model,  fluctuations in the system  propagate spatially in a spin-direction-dependent manner, but the spins themselves are not moving. Like the Malthusian flock, which has no appreciable density fluctuations because the Malthusian balance between birth and death suppresses them, such a spin system clearly has no density fluctuations either, in this case because there is no motion. Therefore, the only hydrodynamic variable in such a spin system is the spin field $\bs(\br,t)$. 
Furthermore, since 
both the symmetries and conservation laws of such a spin system are the same as those of the easy-plane Malthusian flock I consider here,    
the hydrodynamic equation of motion for $\bs(\br,t)$ is exactly the same as the one we derive here for a Malthusian flock, with spin playing the role of the velocity field.

I focus here on the ``polar ordered" state of the Malthusian flock; that is,  a state of an active polar system in which the spatially averaged velocity vector $\langle\bv(\br,t)\rangle$ is non-zero in the thermodynamic limit of $N\to\infty$, where $N$ is the number of active agents (which I will hereafter call ``flockers``). I will choose my coordinate system so that  the direction of the mean velocity is $x$; hence, $y$ is the direction perpendicular to the mean flock motion in the easy plane.

The large-distance, long-time scaling behavior of the easy plane Malthusian flock is characterized by 
three universal exponents: a ``dynamical`` exponent $z$, which relates time scales to length scales along $y$; an ``anisotropy`` exponent $\zeta$, which relates length scales 
along both  $x$ and $r_h$
to length scales along $y$, and a ``roughness`` exponent $\chi$, which relates the scale of velocity fluctuations to length scales along $y$. 

More precisely, the anisotropy exponent $\zeta$,  relates length scales 
along both $y$ and $r_h$ in a ``boosted`` frame moving along the direction of flock motion at a non-universal speed $\gamma$ which is {\it not}, in general, the average speed $v_0$ of the flock itself, although it is likely to be of the same order of magnitude.

 The fact that the same anisotropy exponent $\zeta$ characterizes both the 
$x$ and $r_h$ directions arises from the fact that fluctuations of the velocity
in either direction are ``massive``, in the sense that they will quickly relax back to zero, even if they occur at very long wavelengths. The physical reasons for this are very different for the two directions: fluctuations of the vleocity along $x$ are massive because $x$ is the broken symmetry direction, while fluctuations along $r_h$ are massive because of the explicit rotational symmetry breaking of the underlying dynamics. Despite this difference in the {\it origin} of the ``mass``, the {\it effect} of the mass is the same in both cases: it leads to the same anisotropy exponent for both the $x$ and the ``hard`` ($r_h$) directions, as the RG calculation I`ll present below shows.

That is, the exponents are defined via the ``typical`` time $t(y)$,  ``typical`` perpendicular distance $R_{_\perp}(y)$, and ``typical`` velocity fluctuation $v_y(y)$
\beq
t(y)\propto |y|^z \sep R_{_\perp}(y)\propto |y|^\zeta \sep v_y(y)\propto |y|^\chi \,.
\label{scaledef}
\eeq
where  
\beq
R_{_\perp}=\sqrt{(x-\gamma t)^2+\alpha r_h^2}
\label{rpdef}
\eeq
with $\alpha$  a dimensionless, non-universal (i.e., system-dependent) parameter of $O(1)$.

These exponents can be determined from the correlations of the velocity field $\bv(\br,t)$. I find that the two point velocity correlation functions take the scaling form
\bew
 \begin{eqnarray}
\langle \bv(\br+\br^\prime,t+T)\cdot\bv({\mathbf{r}^\prime,T})\rangle=|y|^{2\chi}F\left({(R_{_\perp}/ a_{_\perp})\over(|y|/a_y)^{\zeta}},{(|t|/\tau)\over(|y|/a_y^z)}\right)
\propto&\left\{
\begin{array}{ll}
|y|^{2\chi}\,,&\left({R_{_\perp}\over a_{_\perp}}\right)\ll \left({|y|\over a_y}\right)^{\zeta},{ |t|\over\tau}\ll  \left({|y|\over a_y}\right)^z\\
R_{_\perp}^{2\chi\over\zeta},&\left({R_{_\perp}\over a_{_\perp}}\right)\gg \left({|y|\over a_y}\right)^{\zeta}, { |t|\over\tau}\ll  \left({R_{_\perp}\over a_{_\perp}}\right)^{z/\zeta}\\\
|t|^{2\chi\over z},&{ |t|\over\tau}\gg  \left({R_{_\perp}\over a_{_\perp}}\right)^{z/\zeta}, { |t|\over\tau}\gg  \left({|y|\over a_y}\right)^z \,,
\end{array}
\right.
\label{scale1}
\end{eqnarray}
\ew
where  $\br=(x,y,r_h)$, $F\left({(R_{_\perp}/ a_{_\perp})\over(|y|/a_y)^{\zeta}},{(|t|/\tau)\over(|y|/a_y^z)}\right)$ is  a scaling function, which is universal up to a non-universal overall multiplicative constant,  $a_{y,\perp}$ are non-universal microscopic lengths, and $\tau$ an equally non-universal microscopic time.

 Scaling laws similar in form to \rf{scale1} have been obtained for many flocking systems\cite{book, Active1, Active2, TT1, TT3, birdrev, toner_prl12, maltd>3prl, maltd>3pre}. For incompressible systems\cite{incomp}, it has proven possible to go further, and actually determine {\it exact} values for these universal exponents.
However, incompressible systems
 are an extreme - indeed, a singular, and unstable under renormalization -  limit of one of the parameters: namely, the limit of inverse compressibility going to infinity.  In contrast, all of the parameters in the theory of Malthusian flocks are finite, except for those that are driven to infinity by renormalization.
 
 The claims made in \cite{TT1,TT3} and \cite{birdrev} for compressible non-Malthusian flocks were shown in \cite{rean} to be incorrect, while similar claims for two-dimensional Malthusian flocks\cite{toner_prl12} are incorrect due to the relevance of a cubic vertex in that problem, similar to the one I will discuss below (which proves in the problem I consider here to be {\it irrelevant}, which is why I can get exact exponents here).

Both references \cite{sc} and \cite{cf} claim to obtain exact exponents for the immortal flocks problem; the former only in $d=2$; the latter in all $d$. The answers claimed in these two references are slightly different. I believe that \cite{sc} is incorrect, since it assumes (incorrectly, in my view) that the time derivative of the Goldstone mode must be a total divergence, which is not the case. Numerous counter-examples to this claim exist: the KPZ equation\cite{KPZ} for one, and the hydrodynamics of crystalline solids\cite{mpp}, smectics\cite{mpp}, and discotic phases\cite{discodyn} for three others. \cite{cf} is more promising, but still depends on a simplification of the full hydrodynamic equations that has not yet been completely justified. It also has difficulties in spatial dimensions $d=2$.

In this paper, I show that, 
despite the fact that this problem proves to be fundamentally non-linear, and in contrast to both ``immortal`` flocks\cite{rean}, and {\it isotropic}  Malthusian flocks\cite{toner_prl12, Malthus_1789, maltd>3prl, maltd>3pre}, I can, 
 thanks to the symmetries of this {\it anisotropic} problem, and the structure of its equation of motion,  determine these scaling exponents 
{\it exactly}\cite{incomp}.  I find:
\beq
z =3/ 2 \sep
\zeta  = 3/4 \sep
\chi = -1/2 \, .
\label{exp}
\eeq
Inserting these results into \rf{scale1} gives 
\bew
\begin{eqnarray}
\langle \bv(\br+\br^\prime,t+T)\cdot\bv({\mathbf{r}^\prime,T})\rangle=|y|^{-1}F\left({(R_{_\perp}/ a_{_\perp})\over(|y|/a_y)^{3/4}},{(|t|/\tau)\over(|y|/a_y^{3/2})}\right)
\propto&\left\{
\begin{array}{ll}
y^{-1}\,,&\left({R_{_\perp}\over a_{_\perp}}\right)\ll \left({y\over a_y}\right)^{3/4},{ |t|\over\tau}\ll  \left({y\over a_y}\right)^{3/2}\\
R_{_\perp}^{-(4/3)},&\left({R_{_\perp}\over a_{_\perp}}\right)^{3/4}\gg{y\over a_y}, { |t|\over\tau}\ll  \left({R_{_\perp}\over a_{_\perp}}\right)^{2}\\\
|t|^{-(2/3)},&{ |t|\over\tau}\gg  \left({R_{_\perp}\over a_{_\perp}}\right)^{2}, { |t|\over\tau}\gg  \left({y\over a_y}\right)^{3/4} \,,
\end{array}
\right.
\label{scale2}
\end{eqnarray}
\ew

 The above results assume the existence of an ordered state. While it is clear that the ordered state is {\it locally} stable, even in the presence of noise (as indicated by the negative value of the roughness exponent $\chi$), recent simulation studies of a spin model that should belong to the universality class of an {\it isotropic} 2d Malthusian flock have found that model to be only metastable: after a very long nucleation time, grain-boundary like defects appear which destroy the ordered state
\cite{besse}. It is unclear whether this  ``meta-instability`` is universal to all Malthusian flocking systems, and even less obvious that it occurs in the easy-plane model I consider here (or, for that matter, in any 3d Malthusian system). Nonetheless, this is obviously a concern. However, even if this meta-instability {\it does} ocur in the system I consider here, it seems likely that the nucleation time for the defects will be long enough to allow the scaling laws I predict here to be tested, since this could be done for the system studied by \cite{besse}.

The remainder of this paper is organized as follows:
in section \rf{eom}, I derive the hydrodynamic equation of motion for easy-plane Malthusian flocks. In section \rf{drg} I use a dynamical renormalization group (DRG) 
analysis of a restricted version of this model to obtain the exact scaling exponents \rf{exp}. In section \rf{irr}, I prove that the terms neglected in section \rf{drg} are, in fact, irrelevant in the renormalization group sense, which implies that the exponents found for the estricted model of \rf{drg} in fact apply in general. Finally, in \rf{sum}, I summarize my results.

\section{Equation of motion}{\label{eom}}

The hydrodynamic equation of motion for a Malthusian flock in an {\it isotropic} space was derived in \cite{toner_prl12}. The hydrodynamic equation of motion I derive here is essentially the same, except for the confinement of the velocity field $\bv(\br,t)$ to lie in the $(x,y)$ plane. 

Here I consider this confinement to be absolutely rigorous: in my approach, the only non-zero components of $\bv(\br,t)$ are 
the $x$ and $y$ components $v_x$ and $v_y$. 

One could imagine relaxing this constraint, and considering anisotropic flocks in which the flockers were allowed to move in the hard direction, but with motion along that direction suppressed by, e.g., decaying with a finite lifetime. However, it is straightforward to show that such a suppression will only grow upon renormalization, leading to complete suppression of the hard component of the velocity as the system is renormalized. Physically, this simply amounts to the statement that, if we look at the system on time scales much longer than the relaxation time of the hard component of the velocity, that component will be effectively zero.
I therefore choose to restrict the velocity field $\bv$ in the hydrodynamic theory to lie in the $(x,y)$ plane.

Similar considerations apply to the {\it magnitude}  $v\equiv|\bv|$ of the velocity 
$|\bv|$. That is, I {\it could} choose to allow this magnitude to fluctuate, as was done in, e.g., the ``Toner-Tu`` equations\cite{TT1,TT3}.  However, since we expect any flocking system to have a preferred speed $v_0$, with fluctuations away from this speed decaying with a finite lifetime, we can instead choose to work from the outset with a velocity which is fixed in magnitude at $v_0$.\cite{polar, nlsig, thesis}  

In addition to the above considerations, I now apply  the general principles of hydrodynamics:

\noindent I) The hydrodynamic variables are the ``slow`` variables: that is, variables that evolve slowly at long wavelengths. There are generically two reasons a variable will be slow in this sense: first, if it is a ``Goldstone mode`` associated with a spontaneously broken continuous symmetry, and second, if it is a conserved variable. 

If we define our $x$-axis to be the direction of the partially averaged velocity 
$\langle\bv\rangle$ of the flock, then the component $v_y(\br, t)$ of velocity perpendicular to that direction is a Goldstone mode, and, hence, will be one of our hydrodynamic variables.

In flocks without birth and death, in which, therefore, the number of flockers is conserved, the local density $\rho(\br,t)$
of the flockers is a slow variable for this reason. This additional variable complicates the hydrodynamics considerably. This greater complexity makes it impossible to determine the exact scaling exponents for such ``immortal`` flocks\cite{rean, sc, cf,  mpp, KPZ, discodyn}. Here, this difficulty does not arise, since $\rho(\br,t)$ is not conserved, and, hence, not a hydrodynamic mode.

Therefore, the only hydrodynamic variable is the velocity field $\bv(\br,t)$. Its equation of motion is constrained by rotational invariance in the $x$-$y$ plane to be\cite{lambda2}:
\bew
\begin{eqnarray}
&&\partial_{t}
v_k(\br,t)=
T_{ki}\bigg(-\lambda (\bv\cdot\vnab)v_i+
   \mu_{_{BE}} \pp^E_i
(\vnab
\cdot \bv) + \mu_{_{TE}}\nabla_{_E}^{2}v_i +\mu_h\pp_h^{2}v_i +
\mu_{_{A}}(\bv\cdot\vnab)^{2}v_i+f _i\bigg)
\label{vEOM}
\end{eqnarray}
\ew
where the indices $i$ and $j$ run over $(x,y)$, and I have defined the transverse projection operator
\beq
T_{ki}\equiv\delta^E_{ki}-v_k(\br,t)v_i(\br,t)/v_0^2
\label{Tdef}
\eeq
which projects any vector orthogonal to $\bv(\br,t)$. Its presence in \rf{vEOM} insures that the fixed length condition $|\bv(\br,t)|=v_0$ on $\bv(\br,t)$ is preserved.


The $\bof$ term in (\ref{vEOM}) is a random Gaussian white noise, reflecting  errors made by the flockers, with correlations:
\begin{eqnarray}
\la f_{i}(\br,t)f_{j}(\br',t')\ra
=2D\delta^E_{ij}\delta^{d}(\br-\br')\delta(t-t')
\label{white noise}
\end{eqnarray}
where the noise strength $D$ is a constant hydrodynamic parameter
and $i , j$ label vector components.

In the ordered state (defined as the state   in which $\left<\bv (\br, t) \right>= v_0 {\bf \hat x}\ne{\bf 0}$, where I`ve chosen the spontaneously picked direction of mean flock motion as our $x$-axis),
we can
write the velocity as:
\begin{eqnarray}
\bv (\br, t) = \sqrt{v_0^2-v_y^2(\br,t)} \, {\bf \hat x} +v_y(\br,t) {\bf \hat y} \,.
\label{vexp}
\end{eqnarray}

Taking the $y$-component of the equation of motion \rf{vEOM},
changing co-ordinates to a new Galilean frame $\br'$ moving with respect to our original frame
in the direction ${\bf \hat{x}}$ of mean flock motion at speed $\gamma=\lambda v_0$ -- i.e., $\br'\equiv\br-\gamma t  {\bf \hat x}$, and, finally, expanding to cubic order in $v_y$ and dropping terms of order $v_y^2$ with more than one spatial derivative,   gives
\bew
\beqn
\pp_t v_y +{\lambda\over2} \pp_y(v_y^2) +\lambda_3\pp_x(v_y^3) &=& \bigg(\mu_x \pp_x^2 + \mu_y\pp_y^2+\mu_h\pp_h^2\bigg)v_y +f_y\ ,
\label{cub}
\eeqn
\ew
where I have defined $\lambda_3\equiv-{\lambda\over6v_0}$, and the diffusion constants $\mu_{x,y}$ are given by
\beq
\mu_x=\mu_{_{TE}}+\mu_{_{A}}v_0^{2}
\sep
\mu_y= \mu_{_{BE}} + \mu_{_{TE}} \,.
\label{mudef}
\eeq
Stability of the homogeneous ordered state requires that $\mu_{x,y,r_h}$  are all positive.

I will begin by studying the equation obtained from \rf{cub} by dropping the cubic 
$\lambda_3$ term, which reads:
\beqn
\pp_t v_y +{\lambda\over2} \pp_y(v_y^2) &=& \bigg(\mu_x \pp_x^2 + \mu_y\pp_y^2+\mu_h^2\pp_h^2\bigg)v_y +f_y \,.
\nonumber\\
\label{nocub}
\eeqn

\section{Dynamic renormalization group (DRG) analysis}{\label{drg}}
 
I now perform a DRG analysis \cite{FNS} on the equation of motion (\ref{nocub}). Specifically, I  first average  over short wavelength degrees of freedom, and then perform the following rescaling:
\beq
x \mapsto  \ee^{\zeta_x\ell}x
\ , \
 r_h\mapsto  \ee^{\zeta_h\ell}r_h
\ , \
y \mapsto  \ee^{\ell} y
\ , \
t \mapsto  \ee^{z \ell} t
\ , \
v_y \mapsto  \ee^{\chi \ell} v_y
\ .
\eeq

There are a number of features of the structure of  the reduced equation of motion \rf{nocub} that prevent certain parameters appearing in \rf{nocub} from being renormalized by the first step of the DRG process (that is, the averaging over the short-wavelength degrees of freedom). In the jargon of the field, these parameters get no ``graphical" renormalization.

The first such parameter is the coefficient $\lambda$ of the only nonlinear term in the equation of motion, $\lambda \pp_y(v_y^2)$. This gets no graphical renormalization because of the inherent pseudo-Galilean symmetry of \rf{nocub}, i.e., the invariance of the equation of motion under the simultaneous
replacements: $y \mapsto y +t\lambda w$ and $v_y \mapsto v_y+w$ for any arbitrary constant $w$.

The parameters $\mu_{x,h}$ also get no renormalization, because that sole non-linearity $\lambda \pp_y(v_y^2)$ is a total $y$-derivative, and, hence, can only renormalize terms in the equation of motion that are also total $y$-derivatives, which the $\mu_{x,h}$ are clearly not.

Finally, the noise strength $D$ gets no graphical renormalization for essentially the same reason: it also involves no $y$-derivatives.

In light of these considerations, the recursion relations for the aforementioned parameters can be obtained purely from the rescaling step of the DRG, and are easily shown to read:
\beqn
\label{eq:D}
\frac{1}{D}\frac{\dd D}{\dd \ell}&=&
z-2\chi -\zeta_x-\zeta_h-1  \,,
\\
\label{eq:l}
\frac{1}{\lambda}\frac{\dd \lambda}{\dd \ell}&=&
z+\chi-1 \,,
\\
\label{eq:mux}
\frac{1}{\mu_{x}}\frac{\dd \mu_{x}}{\dd \ell}&=&
z-2\zeta_x
\,,
\\
\label{eq:muh}
\frac{1}{\mu_{h}}\frac{\dd \mu_{h}}{\dd \ell}&=&
z-2\zeta_h
\,,
\eeqn
{\it exactly}.

To obtain a DRG fixed point,  all  four of these derivatives must vanish. This  gives us  four independent linear equations in the  four  unknown
exponents $z$, $\zeta_x$, $\zeta_h$, and $\chi$, whose solutions are  
\beq
\zeta_h=\zeta_x\equiv\zeta=3/4 \sep z=3/2 \sep \chi=-1/2 \,.
\label{scexp}
\eeq

With these exponents in hand, a straightforward and entirely standard DRG analysis implies that the two-point velocity-velocity correlation function has the form \rf{scale2},  with $\zeta$ being defined, as in \rf{scexp}, as the common value of $\zeta_h$ and $\zeta_x$.

\section{Proving the irrelvance of $\lambda_3$} {\label{irr}}

Since both the $\lambda$ and the $\lambda_3$ vertex in equation \rf{cub} actually originate from the same rotation-invariant $(\bv\cdot\nabla)\bv$ term in our original equation of motion \rf{vEOM}, rotational invariance implies that the ratio of these two term remains fixed at $-6v_0$. Since our system has long-ranged orientational order, which is equivalent to a non-zero average velocity, $v_0$  is asymptotically unrenormalized at large distances\cite{finite}. Thus, the ratio of $\lambda$ to $\lambda_3$ must remain unchanged by the graphical corrections to them. Therefore, the DRG recursion relations for these two parameters must read:

\beqn
\frac{d\ln \lambda}{d\ell}&&=z+\chi-1+f(\lambda_3, ...) \,,
\label{l2}
\nn\\
\frac{d\ln \lambda_3}{d\ell}&&=z+2\chi-\zeta+f(\lambda_3, ...)\,,
\label{l3}
\eeqn
where $f(\lambda_3, ...)$ denotes the graphical corrections.\cite{AS}

My earlier argument that $\lambda$ gets no graphical correction when $\lambda_3=0$ implies that 
$f(\lambda_3, ...)$ vanishes as $\lambda_3\to0$. Therefore, it is at most $O(\lambda_3)$ for small $\lambda_3$. Hence the recursion relation \rf{l3} for 
$\lambda_3$ implies
\beq
\frac{d \lambda_3}{d\ell}=(z+2\chi-\zeta)\lambda_3+O(\lambda_3^2)=-{1\over4}\lambda_3 +O(\lambda_3^2)\,.
\label{l3.2}
\eeq
Thus I see that a sufficiently small $\lambda_3$ is always irrelevant, with RG eigenvalue $-1/4$.  This justifies dropping the $\lambda_3$ term, as I did earlier.

\vspace{.2in}

\section{Summary}{\label{sum}}

I have formulated the hydrodynamic theory of three dimensional ``Malthusian flocks" moving with an easy plane. This problem proves to be one of the rare problems\cite{incomp} for which, despite strongly relevant non-linearities in the equation of motion, it is possible to determine the universal scaling exponents characterizing it {\it exactly}, finding that the dynamical exponent $z=3/2$, the ``anisotropy" exponent $\zeta=3/4$, and the ``roughness" exponent $\chi=-1/2$. I`ve also determined  the scaling form of the velocity-velocity two point correlation function implied by these exponents.

\begin{acknowledgments}
I am grateful to Sriram Ramaswamy and Ananyo Maitra for extremely valuable discussions, without which I would not have been able to demonstrate the irrelevance of the cubic vertex $\lambda_3$, on which my entire argument for exact exponents depends.  I also  thank  The Higgs Centre for Theoretical Physics at the University of Edinburgh, where this work was done,  for their hospitality and support.
\end{acknowledgments}


\begin{thebibliography}{7}
\bibitem{polar} Alternatively, one could instead choose to take as the hydrodynamic variable in the problem the unit vector $\hp\equiv\bv/|\bv|$ along the velocity. This variable, often called ``the polarization``, is, indeed, often used in the field of Active Matter. I choose not to do so here. The difference between these choices is irrelevant in both the renormalization group and the conventional English sense.   This is widely accepted in the field of critical phenomena; it is the reason, for example, that the non-linear sigma model\cite{nlsig1, nlsig2, nlsig3, pn} is believed to belong to the same universality class as the Landau-Ginsburg-Wilson theory of ferromagnets.
One way it can be seen in the context of the present work is that the ultimate hydrodynamic equations for the Goldstone modes that are pbtained in \cite{rean} are exactly those that one obtains by starting with a fixed-length model as I have done here. The choice of whether to use a fixed-length or soft spin model is therefore purely a matter of convenience.
\bibitem{lambda2} Readers familiar with the ``Toner-Tu`` equations might wonder why there is no term proportional to $\lambda_2(\nabla\cdot\bv)v_i$ in this equation of motion. The reason is that such a term is along the velocity, and so vanishes when projected perpendicular to the velocity. That is, if we were to include the $\lambda_2$ term in the equation of motion \rf{vEOM}, it would read
\bew
\begin{eqnarray}
&&\partial_{t}
v_k(\br,t)=
T_{ki}\bigg(\lambda_2(\nabla\cdot\bv)v_i-\lambda (\bv\cdot\vnab)v_i+
   \mu_{_{BE}} \pp^E_i
(\vnab
\cdot \bv) + \mu_{_{TE}}\nabla_{_E}^{2}v_i +\mu_h\pp_h^{2}v_i +
\mu_{_{A}}(\bv\cdot\vnab)^{2}v_i+f _i\bigg) \,.
\label{vEOMl2}
\end{eqnarray}
\ew
But 
\beq
T_{ki}\lambda_2(\nabla\cdot\bv)v_i=\lambda_2(\nabla\cdot\bv)T_{ki}v_i=0 \,,
\eeq
since $T_{ki}v_i=0$ by construction (that is, since $T_{ki}$ projects orthogonal to $\bv$). Hence, the $\lambda_2$ term drops out of the equation of motion.
In reference \cite{rean}, it was shown that this also holds true (that is, that the $\lambda_2$ term drops out of the final equation of motion for the Goldstone mode)  for  ``soft spin`` flocks, in which the magnitude of the speed is allowed to fluctuate.
\bibitem{finite}  There are, of course, {\it finite} fluctuation renormalizations of the speed $v_0$. However, since these remain finite as system size $L\to\infty$, it must be the case that ${dv_0\over d\ell}\to0$ as renormalization group time $\ell\to\infty$. Hence, when finding an DRG fixed point, which, of course, is only {\it asymptotically} approached in the limit of large DRG time $\ell$, we can treat $v_0$ as a constant under the DRG.
\bibitem{AS} This aproach to determining the relevance of $\lambda_3$,  was first obtained, for the somewhat different problem of a two-dimensional Malthusian flock, by Sriram Ramaswamy and Ananyo Maitra. I thank them for patiently explaining this argument to me.
\end{thebibliography}
\end{document}